\documentclass[aps,prb,twocolumn,superscriptaddress]{revtex4-2}
\usepackage{amsmath}
\usepackage{amssymb}
\usepackage{color}
\usepackage{feynmp-auto}
\usepackage{graphicx}
\usepackage{grffile}
\graphicspath{{images/}}
\graphicspath{{figures_prb/}}
\begin{document}

\title{Local density of states of the interacting resonant level model at zero temperature }

\author{G.~Camacho}
\affiliation{Technische Universit\"at Braunschweig, Institut f\"ur Mathematische Physik, Mendelssohnstrasse 3, 38106 Braunschweig, Germany}

\author{P.~Schmitteckert}
\affiliation{HQS Quantum Simulations GmbH,
Haid-und-Neu-Stra{\ss}e 7,
76131 Karlsruhe, Germany}

\author{S.~T.~Carr}
\affiliation{School of Physical Sciences, University of Kent, CT2 7NH Canterbury, United Kingdom}

\setlength{\abovedisplayskip}{4pt}
\setlength{\belowdisplayskip}{4pt}

\date{\today}

\begin{abstract}
The Anderson single impurity model, including its Kondo limit, can be seen as the hydrogen atom for correlated systems. Its most striking feature is the universal scaling of its low energy properties governed by the appearance of a single emergent scale, the Kondo temperature $T_{K}$. In this work, we demonstrate the emergence of a second independent energy scale in a quantum impurity model, the interacting resonant level model, which is equivalent to the anisotropic Kondo model. We study the local density of states at the impurity site in the ground state (zero temperature limit) of the model. We see collapse of data onto universal curves at low energies, however this is only achieved by defining a second low energy scale with a different power law dependence on model parameters than the Kondo temperature. We provide an exact expression for this second critical exponent. We also report on a splitting of the central resonance as the interaction strength is increased in the absence of any external magnetic field in the model.\\
\end{abstract}

\maketitle


\section{Introduction}
   Since the times of Kondo~\cite{Kondopaper} and Anderson~\cite{Andersonmodel}, physicists have been fascinated by the possible effects of embedding a correlated quantum impurity in a metallic host~\cite{Affleck_lectures,Saleur_lectures,TsvelickWiegmann,Costi}. As a consequence of the interaction between the impurity and the surrounding electrons, an emergent low energy scale appears: in the Anderson and Kondo models, this scale is the Kondo temperature
$T_K$, which dominates all low energy properties of the
system: physical observables in these models show
universal results when scaled against $T_K$. It is however
not clear if such universal scaling paradigm, with $T_K$
as the only relevant low energy scale, is expected to
hold for all observables in quantum, single impurity
models. In the case of multi-orbital impurities or impurity lattice models (like in the Kondo lattice model), an interplay between multi-orbital interactions and Kondo physics is expected to break such single energy picture, leading to the appeareance of distinct energy scales, for instance, when considering spin-spin interactions along with multi-orbital interactions; however this is not expected when only a single impurity is present, specially if spin degrees of freedom are not accounted for. Most known single impurity models have reported the appeareance of a single energy scale, in accordance with the exact solutions found by Bethe ansatz~\cite{Filyov1980,AndreiKondo,Wiegmann1981,TsvelickWiegmann,Rylands}. While static ground state properties like
the impurity susceptibility are expected to follow such
scaling, observables depending on excited states of the
spectrum are subjected to strong correlation effects that
can potentially break the universal picture. Yet,
these strong correlations can ultimately lead to the
appeareance of additional low energy scales that restore
universal results. 
  
Recently, a lot of work in the area of quantum impurity models has concentrated around the Interacting Resonant Level model (IRLM)~\cite{WiegmannFinkelshtein,Doyon,andergassen,Kiss2,borda_perturbative_irlm,Boulat1,Andrei,Samfcs,Boulat2,Kennes_quench,SaleurIRLMBSG,Sorantin,BBSS,SchillerAndrei,Karr10,Borda1}; first introduced in 1978~\cite{WiegmannFinkelshtein} as formally equivalent to the anisotropic Kondo model, it gained a lot of interest in its own right since various exact solutions out of equilibrium were proposed for it~\cite{Filyov1980, Andrei, Boulat2, Samfcs,Braun}. Despite a lot of progress being made on non-equilibrium transport in the IRLM \cite{Doyon,andergassen,Kiss2,borda_perturbative_irlm,Boulat1,Boulat2,BBSS,Andrei,
Samfcs,Borda1,SchillerAndrei,Karr10,
SaleurIRLMBSG,Kennes_quench,Sorantin}, there remains open questions about equilibrium properties. 

Previous studies in the IRLM~\cite{WiegmannFinkelshtein,Kiss1,Borda1,Borda2,ohmic_model,Schlottman} have shown that certain ground state properties such as the local susceptibility depend on a single energy scale, the equivalent of the Kondo temperature. This energy scale has a power law dependence on the hybridisation between the impurity and the leads; in a recent work~\cite{paper_thermodynamics} we showed that this power can be calculated exactly, even for strong interactions. We also showed that the shape of this susceptibility as a function of a local field only varies slightly as a function of interaction, through a formula originally given in previous works~\cite{Rylands,Ponomarenko}, with the profile shape being a Lorentzian. In this work, we concentrate on the local density of states (LDoS) of the IRLM in the ground state, which unlike the susceptibility, depends on excited states of the spectrum. While in the absence of interactions these two properties are identical, we will show that in the presence of interactions, the local
density of states depends on an additional (emergent)
energy scale with a different critical exponent than that
of the susceptibility.

\section{Model}
The interacting resonant level model consists of a one-dimensional non-interacting lead of spinless fermions coupled to a single level via a weak hybridisation $t'$ and an interaction $U$. The lead is modeled as a tight binding chain with hopping energy $t$. The Hamiltonian for the single lead IRLM is written in second quantised notation as: 
\begin{multline}\label{IRLMham}
\mathcal{H}=-t\sum_{n=0}^{N-1}c_{n+1}^{\dagger}c_{n}+\text{H.c.} + \varepsilon_{0}d^{\dagger}d + t'(c_{0}^{\dagger}d + \text{H.c.})\nonumber\\
+ U\left( d^{\dagger}d-1/2\right)\left(c_{0}^{\dagger}c_{0}-1/2\right).
\end{multline}
The $c$ and $d$ operators are of fermionic nature: $d$ is the resonant level annihilation operator, and $c_{n}$ are (spinless) fermion annihilation operators at a site $n$ on the lead.
The impurity level is resonant if its energy $\varepsilon_0$ is equal to the Fermi level on the lead, in this case both will be zero.
 
\subsection{Local density of states}
The main object of study in this work is the local density of states at the impurity site, defined by: 
\begin{eqnarray}\label{eq_A}
A(\omega)=\text{Im}\left[\frac{i}{\pi}\int_{0}^{\infty}dte^{i\omega t}\langle d(t)d^{\dagger}(0) + d^{\dagger}(0)d(t)\rangle\right]
\end{eqnarray} 
This quantity is a measure of excitations in the ground state of the system, and may be directly measured in scanning tunneling microscopy experiments~\cite{STM}. In the non-interacting system ($U=0$), this is given by a Lorentzian with $T_{0}=\pi\nu(t')^{2}$ representing the resonance width, with $\nu=1/\pi t$ being the bulk density of states in the wide band limit of the uniform chain. To study the change in the resonance shape as interaction is increased we use the Numerical Renormalisation Group (NRG). For details on the method we refer the reader to the literature review~\cite{Costi}.

\begin{figure}[!t]
\begin{center}
\includegraphics[clip=true,width=3.4in, height=2.2in, scale=0.8]{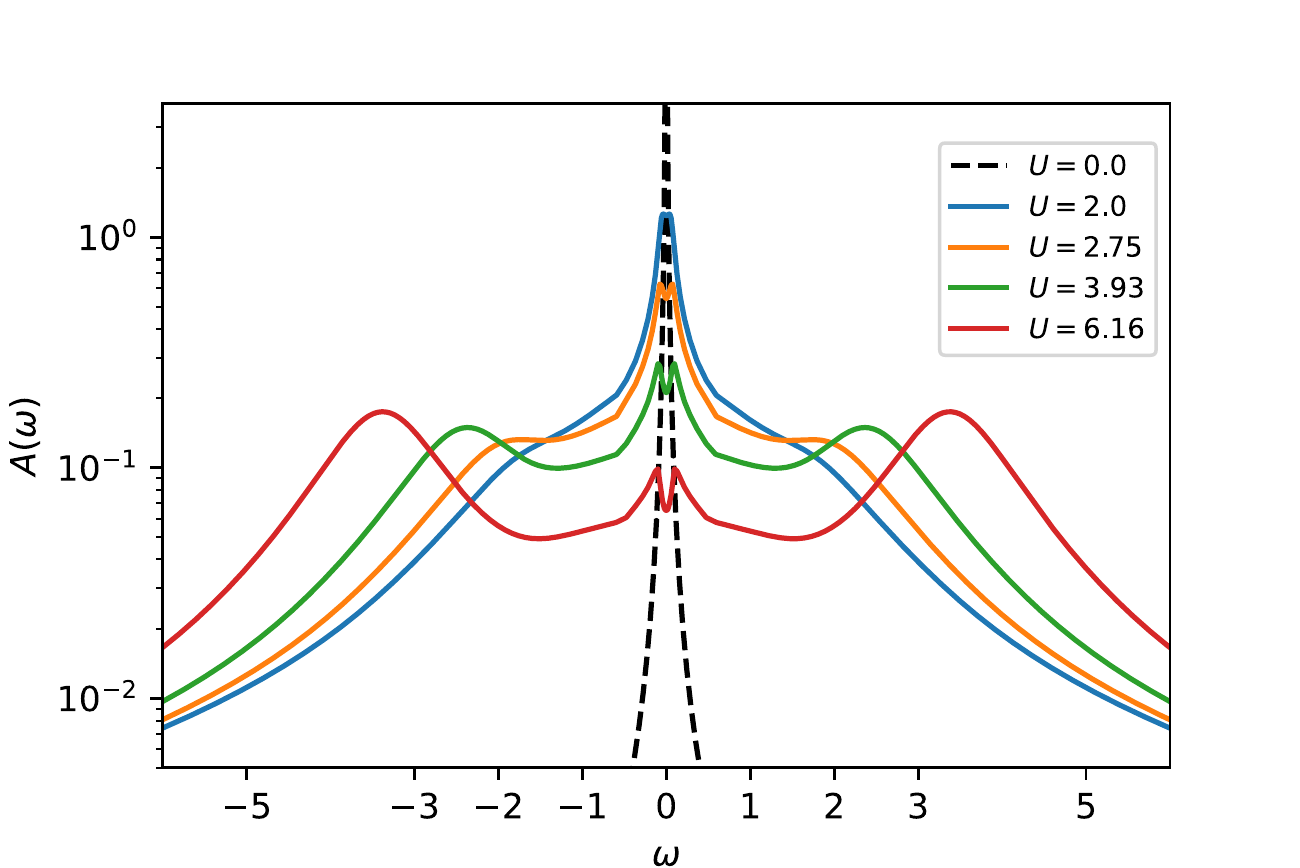}
\end{center}
\caption{Spectral function evolution with interaction $U$ for a fixed value $t'=0.05$, showing the appeareance of Hubbard bands in the high-energy region as spectral weight at lower energies is lost. For the Hubbard bands resolution, standard Lorentzian broadening has been used. NRG data corresponds to $\Lambda=1.3$, with $S_{K}=1000$ states kept after truncation and a system of $N=92$ sites.  }
\label{figure0}
\end{figure}

 In Fig.~\ref{figure0} we plot $A(\omega)$ for different vales of $U$ over a large energy range. The main thing to notice is that the spectral weight shifts from the central resonant peak towards high energies. These side peaks appearing at roughly $\omega=\pm U/2$ are equivalent physics to the Hubbard bands in the Anderson impurity model~\cite{Costi}; similar physics has also been previously seen in a two-lead version of the interacting resonant level model~\cite{Braun}. We now concentrate on the evolution of the low energy resonant peak for $\omega\nu\ll 1$, always keeping in mind that for large interactions, most spectral weight is both in the Hubbard bands and the low-energy sector.


\begin{figure}[!t]
\includegraphics[clip=true,width=3.2in, height=2.3in, scale=0.5]{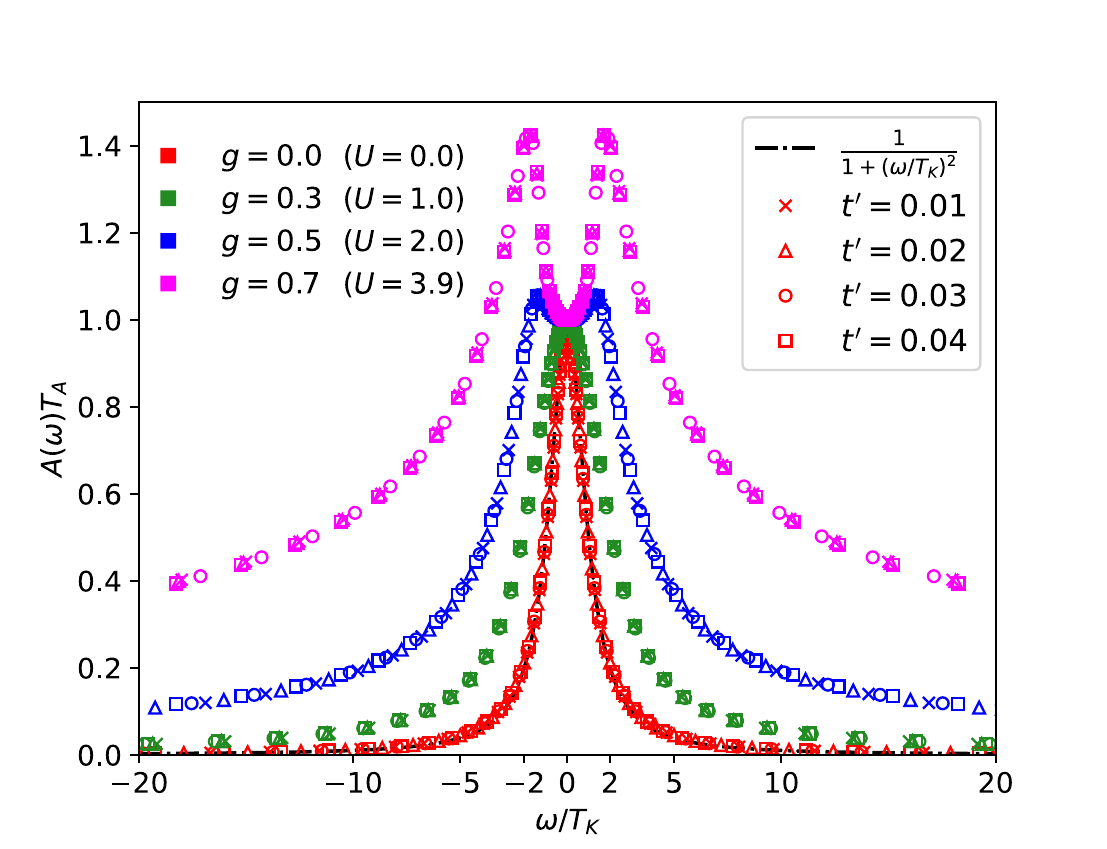}
\caption{Scaled plot for the evolution of the LDoS for different values of $g$ and $t'$ showing \emph{universal} behaviour when both energy scales $T_{A}$ and $T_{K}$ are considered. NRG parameters are $\Lambda=1.5$ with $S_{K}=800$ states kept after truncation, and a system size of $N=92$ sites. }
\label{figure1}
\end{figure}

\begin{figure*}[!t]
\begin{center}
\includegraphics[scale=0.1, height=2.7in, width=1.0\textwidth]{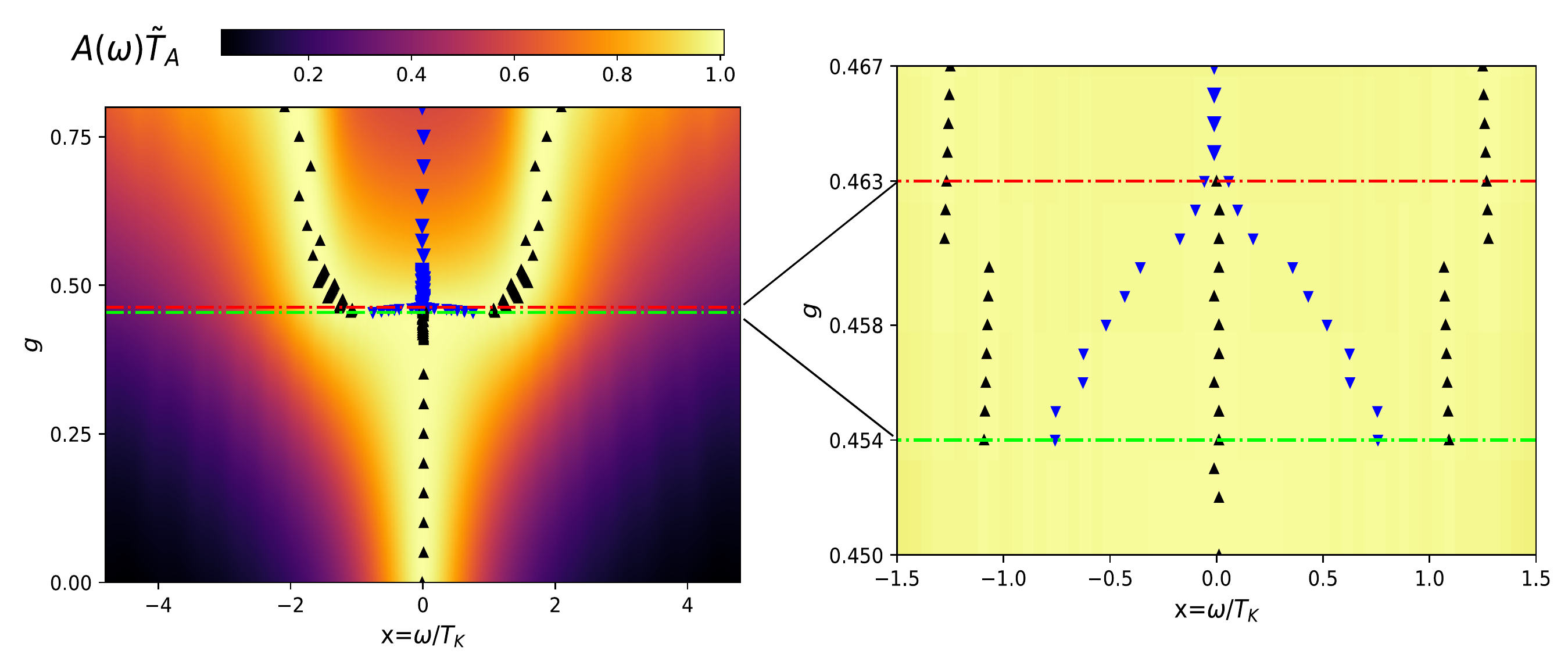}
\end{center}
\caption{ Density plot of the evolution with interaction of the spectral function normalised to its maximum value. Local maxima (black up triangles) and minima (blue down triangles) have been extracted numerically. The green dashed line at $g_{c1}\approx 0.454$ and red dashed line at $g_{c2}\approx 0.463$ separate three different regions: a region for $g<g_{c1}$ where there is a central maximum at $\omega\sim 0$; a region with $g_{c1}<g<g_{c2}$, where the splitting of the central resonance gives two side peaks at $\omega\sim \pm T_{K}$, but maintains the local maxima at $\omega\sim 0$, consequently having two local minima; the $g>g_{c2}$ region, where the central local maxima at $\omega\sim 0$ has turned to a local minima between $\omega=\pm T_{K}$. NRG parameters correspond to $\Lambda=1.2$ with $S_{K}=1200$ and a chain of $N=182$ sites. The discretisation due to the NRG is more apparent in the position of the minima on the right figure. }
\label{figure_2}
\end{figure*} 

\section{Results}
\subsection{Resonant peak}
For an analytical treatment of low energy properties, it is more convenient to use the following dimensionless coupling $g$ rather than the bare interaction $U$:
\begin{eqnarray}\label{g_factor}
g=\frac{2\delta}{\pi}=\frac{2}{\pi}\arctan\bigg(\frac{U\pi\nu}{2}\bigg), && g\in[0,1]\leftrightarrow U\geq 0.
\end{eqnarray}
In this expression, $\delta$ represents the scattering phase shift of fermions at the Fermi level when reaching the lead boundary~\cite{
Borda1,Borda2,Kiss1,paper_thermodynamics}, which is given by the above expression for a tight-binding lead.  For leads with a different band structure or field theories with a different regularisation, the relationship between $g$ and $U$ would change, however the low-energy physics for a given phase shift is universal.

If scaled correctly, the low-energy shape of the spectral function is independent of the hybridisation between the impurity and the leads, $t'$. The resonance width $T_{K}$ (equivalent to the Kondo temperature and known from previous works~\cite{paper_thermodynamics}-- we define it precisely later in Eq.~\eqref{scales_def}) defines a natural dimensionless variable $\omega/T_K$. We define another energy scale as $T_A = 1/A(\omega=0)$, meaning that the scaled spectral function is always $1$ at zero energy.  If we look at data plotted in Fig.~\ref{figure1}, we see that this scaled quantity $A(\omega)T_{A}$ for different values of $t'$ collapse onto universal curves dependent only on the interaction strength $g$.

We defer analysis on the scaling behavior for the next section; for now we concentrate on the most obvious feature of this plot which is that the (low-energy) central peak splits in two as interaction strength is increased. We stress that this splitting is not related to the Hubbard bands in Fig.~\ref{figure0} which occur at much larger energy -- this is an intrinsic splitting of the low-energy resonant peak. We also emphasise that unlike previous work on the splitting of the resonance peak in the Kondo model~\cite{AKMsplit, Niu, CostiAKM, akm_split}, this occurs without an external field splitting the energies of the resonant level, or coupling to any other degree of freedom. Another aspect that makes this evolution particularly surprising is that it is not seen in the susceptibility $\chi=-\left(\partial \langle d^{\dagger}d\rangle/\partial\varepsilon_{0}\right)$ which remains approximately in Lorenzian shape for all values of interaction~\cite{paper_thermodynamics}.

The physics of this splitting is relatively straightforward to see in the strong interaction $g\rightarrow 1$ ($U\rightarrow \infty$) limit, where the resonant level and the final site of the lattice must contain exactly one electron between them at low-energy to avoid paying the high interaction cost $U$. The effective model consists of a two-level system weakly coupled to the rest of the lead; however this two-level system is not resonant as the energies are split by the hybridisation $t'$ (which is equivalent to $T_K$ in the strong interacting limit~\cite{paper_thermodynamics}). This accounts for the splitting in the strong coupling limit. 

To investigate the splitting numerically, we have plotted the scaled spectral function $A(\omega/T_K,g)\tilde{T}_A(g)$ in Fig.~\ref{figure_2} as a density plot, with $\tilde{T}_A(g)\equiv\text{max}(A(\omega,g))$~\footnote{The scaling $\tilde{T}_A$ is slightly different from $T_A$ in Fig.~\ref{figure1} and is chosen to normalise the peak height to $1$ rather than $A(\omega=0)$ -- this difference is unimportant physically but makes the detail in the figure much easier to see.}. We superimpose on top of the density plot the locations of local minima and maxima as a function of $\omega$ for a given interaction strength.  This shows a somewhat surprising feature -- the central peak doesn't in fact split. Rather, side peaks appear at $\omega \sim \pm T_K$ at some critical interaction strength $g_{c1} \approx 0.454$, with the central peak turning into a local minimum at a slightly higher interaction $g_{c2} \approx 0.463$. 

\begin{figure}[!t]
\begin{center}
\includegraphics[clip=true,width=3.4in, height=2.4in, scale=0.7]{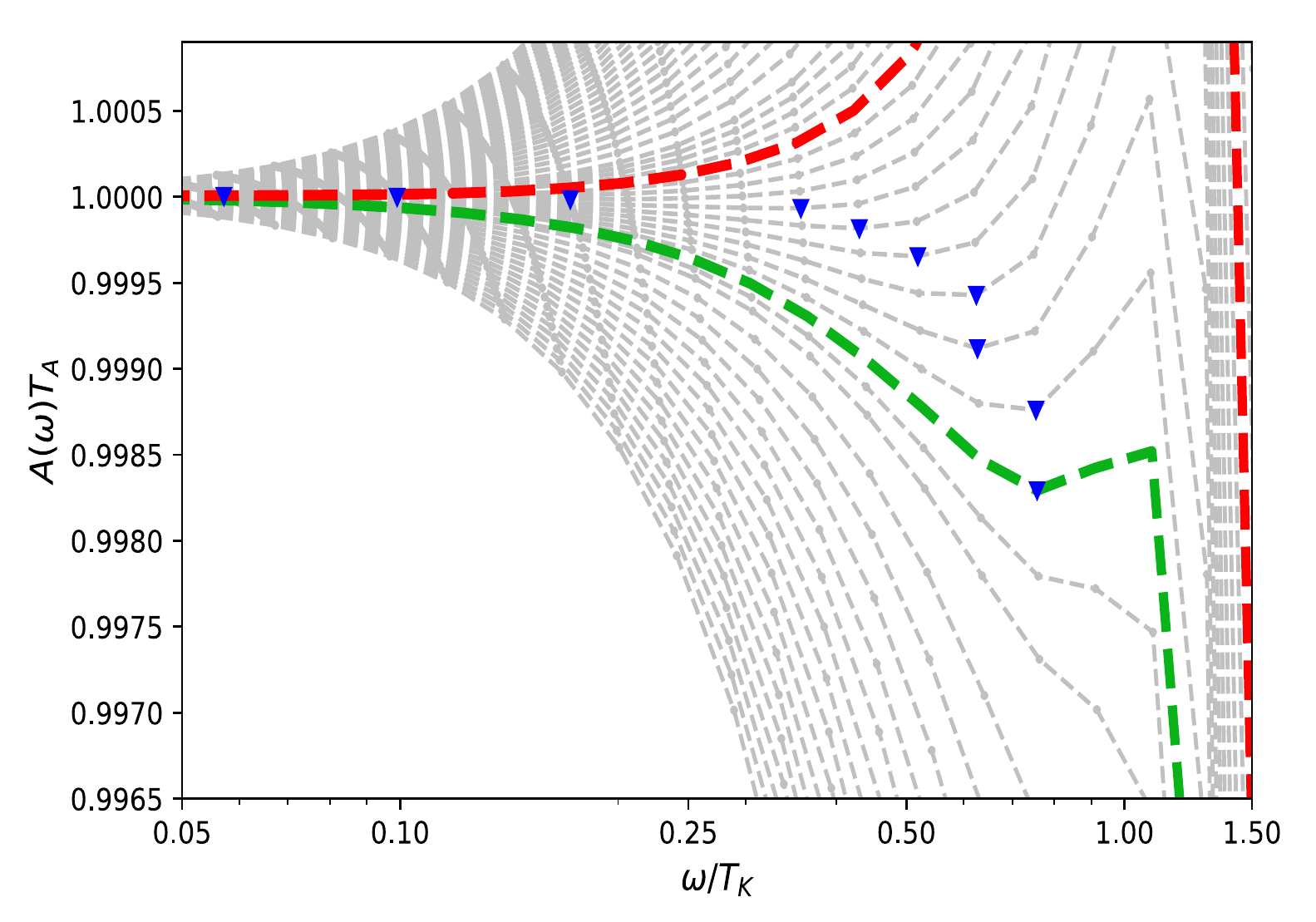}
\end{center}
\caption{A zoom in the $\omega>0$ region  of Fig.~(\ref{figure_2}) where the side peak starts to develop. Notice the change in the curvature with increasing interactions.}
\label{figure4_1}
\end{figure}

This supports a scenario of a general shape of the spectral function consisting of one peak centered on $\omega=0$ with another two centered at $\omega\sim\pm T_K$.  As interaction is increased, the relative weight of the central peak (corresponding to the weak-coupling resonant dot) decreases, while that of the side peak (corresponding to the strong-coupling off-resonant effective dot) increase.  In this sense, the transitions should be thought of more as a crossover, with there being no particular significance to $g_{c1}$ or $g_{c2}$. The detail for the crossover between the weak and strong coupling limits in the $\omega\sim 0$ region is represented in Fig.~\ref{figure4_1}. Over this range of intermediate interactions, the coefficient of the $\omega^2$ term in the spectral function is much smaller than would be expected by the width of the resonance (i.e. the scale $T_K$). It is an open question how this region with relatively incoherent electronic excitations would manifest itself in other observable properties.

\subsection{Energy scales}
We move now to the scaling analysis. In previous works on thermodynamic quantities, it has been observed that $T_K$ is the only low energy scale in the system, with a collapse of data for different values of $t'$ so long as energies are scaled by $T_K$~\cite{paper_thermodynamics,Berkovits2}.  Numerical studies have even shown the single energy scaling for non-equilibrium steady state transport~\cite{Boulat1,Karr10,SaleurIRLMBSG}, sub-leading corrections to full-counting statistics~\cite{Schmitteckert2014,Carr2015}, quenches~\cite{Kennes_quench} and out-of-equilibrium entanglement entropy growth~\cite{Bidzhiev}.

We will show however for the spectral function, one needs to define two distinct energy scales which depend on the hybridisation $t'$ in different ways.  To be precise, these are defined as
\begin{equation}
T_K^{-1} = \chi(\epsilon_0=0),\;\;\;\; T_A^{-1} = A(\omega=0).
\label{scales_def}
\end{equation}
The scale $T_K$ is given by a power law in the hybridisation $T_{K}\sim (t')^{\alpha}$, where $\alpha=2$ for the non-interacting case, while previous work~\cite{paper_thermodynamics} has given an exact expression $\alpha=2/(1+2g-g^2)$ when interactions are present. We use the known expression for $T_K$ from previous works~\footnote{Technically there is also an interaction dependent pre-factor as well as the exponent -- this is slowly varying and is unimportant for scaling behavior, but to be precise, this pre-factor is included to get the $T_K$ used in plots.}, which collapses multiple data sets for different hybridisations $t'$ onto single curves.

\begin{figure}[!t]
\begin{center}
\includegraphics[clip=true,width=3.2in, height=2.6in, scale=0.9]{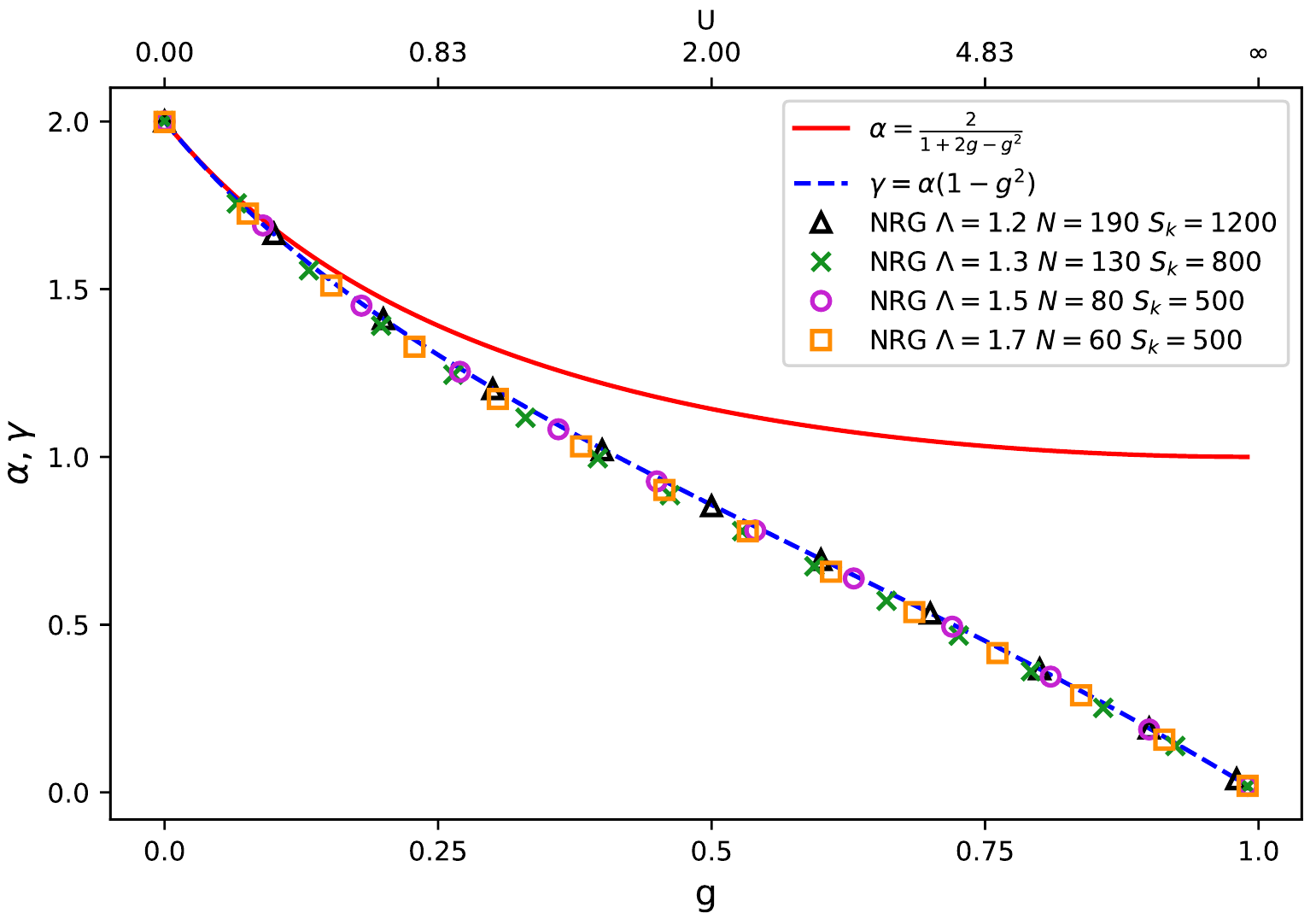}
\end{center}
\caption{The scaling exponent $\gamma$ associated to the energy scale $T_{A}$ Eq.~(\ref{scales_def}) for different NRG parameter sets. The $\alpha$ exponent curve is included for comparison. }
\label{figure4}
\end{figure}

\begin{figure}[!t]
\begin{center}
\includegraphics[clip=true,width=3.2in, height=2.6in, scale=0.9]{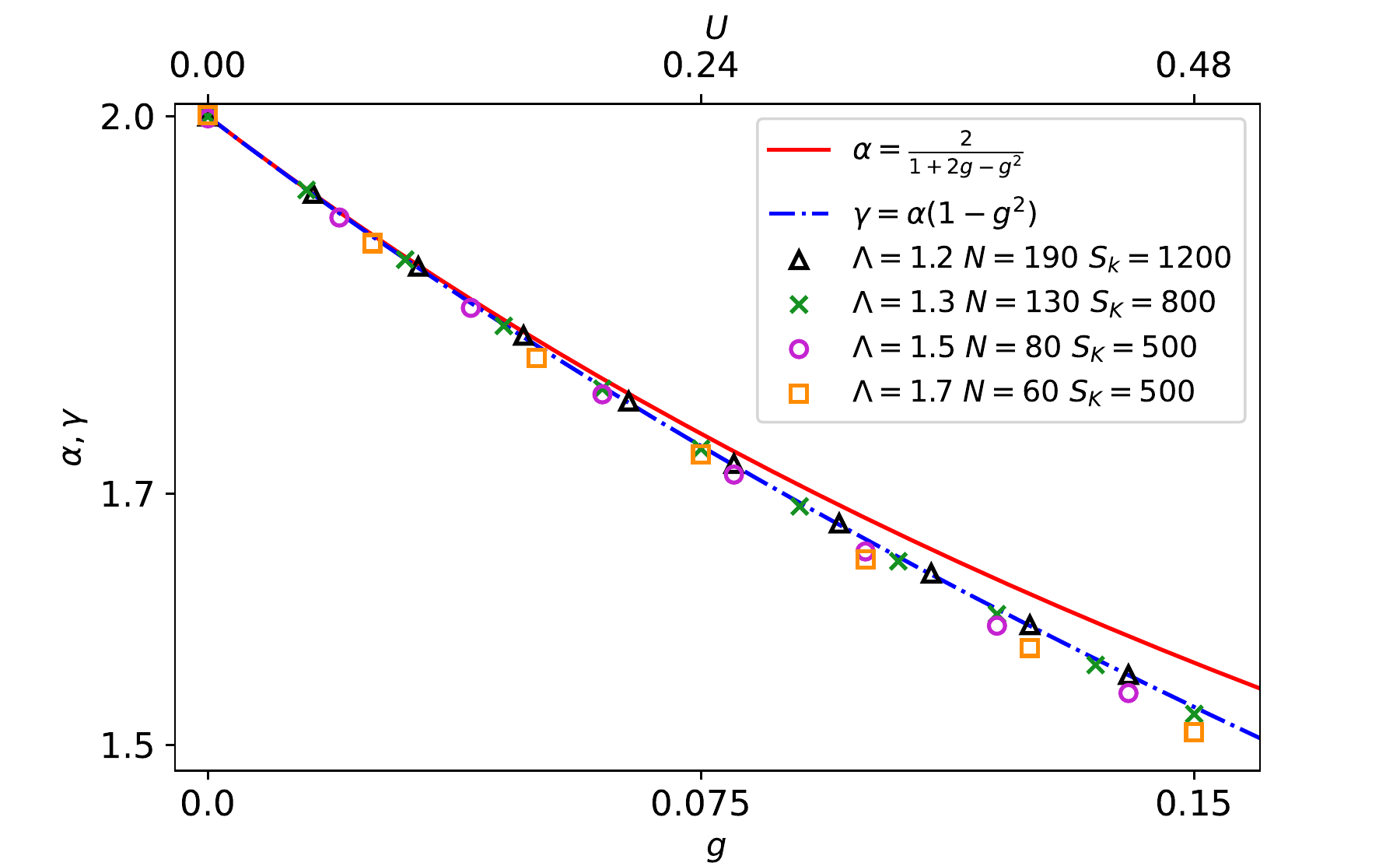}
\end{center}
\caption{ NRG data for the weak coupling sector, showing the deviation of data points from the $\alpha$ curve at ($O(g^{2})$) in the interaction. }.
\label{figure5}
\end{figure}

If one tries to scale the vertical axis with the same $T_K$ however, one does not get collapse.  Defining $T_A$ as in Eq.~\eqref{scales_def}, we can extract this scale from NRG simulations \footnote{We do this by fitting $A(\omega)=T_A+B\omega^2$ for data satisfying $0.01<\omega/T_K<0.1$ as NRG doesn't get all the way to zero energy.} for multiple values of $t'$ and fit $T_A = C(t')^\gamma$ for each value of interaction.  The results are shown in Fig.~\ref{figure4} where  empirically, we can fit
\begin{equation}
\gamma = \alpha(1-g^2) = \frac{2(1-g^2)}{1+2g-g^2}.
\label{gamma_exp}
\end{equation}
In Fig.\ref{figure5}, we show that even for small interactions, we numerically distinguish between the two exponents. 

Let us make some comments here: $T_A$ is an energy scale, and the ratio $T_K/T_A\sim (t')^{\alpha-\gamma}$ can be made arbitrarily small by changing $t'$ for any finite interaction.  In the non-interacting system however the susceptibility is equal to the density of states, $T_A=T_K$ -- and while there is no reason to expect this expression to hold for finite interactions, the fact that they have different exponents indicates that the IRLM is fundamentally a non-perturbative problem~\cite{Saleur2013}. Secondly, there is precedent for the height of a spectral function to scale differently from the width -- indeed, in the Anderson impurity model in the Kondo regime, the Langreth theorem~\cite{Langreth} states that $A(\omega=0)$ is unrenormalised by the on-site interaction $U$.  In this case however, one has the emergent Kondo scale and the original bare energy scale, rather than  two-different emergent low-energy scales.  Finally, we note that multiple exponents have been mentioned before in the IRLM~\cite{Berkovits1}, however these have all been with Luttinger liquid leads -- a second distinct exponent was not expected with non-interacting leads as in this case.

This second exponent~\eqref{gamma_exp} can also be derived analytically, see Appendix \ref{app_B} for details. The derivation relies on a perturbative diagrammatic expansion to second order in interaction~\cite{Schlottman}, however as seen in Fig.~\ref{figure4} it gives a result that appears to be numerically correct for any interaction strength, including the strong interaction regime where the spectral function at low energy seems to bear no resemblance to a Lorentzian shape. It is somewhat surprising that this perturbative method appears to give an exact result even for strong interactions. The exact calculation of the exponent by using non-perturbative techniques such as bosonization is an open question; however, it seems likely that in this case, the equivalence to the sine-Gordon model that works so well for the susceptibility~\cite{paper_thermodynamics} as well as steady-state transport~\cite{SaleurIRLMBSG} can not be applied. 

\section{Summary}
To summarize, we have studied the impurity spectral function of the interacting resonant level model over the full range of interactions from weak to strong.  Our two main results are
\begin{enumerate}
\item As interaction is increased, the spectral function begins to deviate significantly from a Lorenzian shape, eventually splitting into two separate peaks at $\omega\sim T_K$ for interactions $U\nu\sim 1$. This splitting occurs in the absence of any external magnetic field. 
\item The height of the spectral function $A(\omega)$ defines a new energy scale $T_{A}$  which scales with the hybridisation $t'$ with a different exponent than $T_{K}$. The expression for the exponent which is obtained from perturbative RG appears to work for all interaction strengths.

\end{enumerate}
It is an open question whether such features of the spectral function would also manifest in any other observable quantities either in equilibrium or out. Finally, due to the exact mapping between the IRLM and the anisotropic Kondo model~\cite{WiegmannFinkelshtein}, we believe that such features should emerge in the case of strong $S_{z}$ interaction with perturbative $S_{xy}$ exchange. While this is a rather unusual limit to explore in the context of Kondo physics, such results could be observed directly when computing crossed correlations of the spin operator representing the impurity; in particular the ground-state averaged correlator given by:
\begin{eqnarray}\label{Kondo_correlator}
G^{-+}(t)=-i\langle S^{-}(t)S^{+}(0) + S^{+}(0)S^{-}(t)\rangle
\end{eqnarray}
in the anisotropic Kondo model is directly related to the spectral function of the IRLM ( compare Eq.~\eqref{eq_A} ), due to the exact relation existing between the two models~\cite{Saleur_lectures,TsvelickWiegmann,WiegmannFinkelshtein}. An exact bosonization mapping between the two models can also be found in \cite{paper_thermodynamics}. It is an interesting question in which situations this limit of the Kondo model might arise naturally, allowing for experimental realisation of the physics described here.

\begin{acknowledgments}
The authors thank H. Saleur for many fruitful discussions about the IRLM. G.C. was supported by the
Deutsche Forschungsgemeinschaft (KA 3360/2-1) ‘Niedersächsisches
Vorab’ through the ‘Quantum- and Nano-
Metrology (QUANOMET)’ initiative within the project P-1.
\end{acknowledgments}

\appendix
\section{NRG}

\begin{figure}[!h]
\begin{center}
\includegraphics[clip=true,scale=0.8]{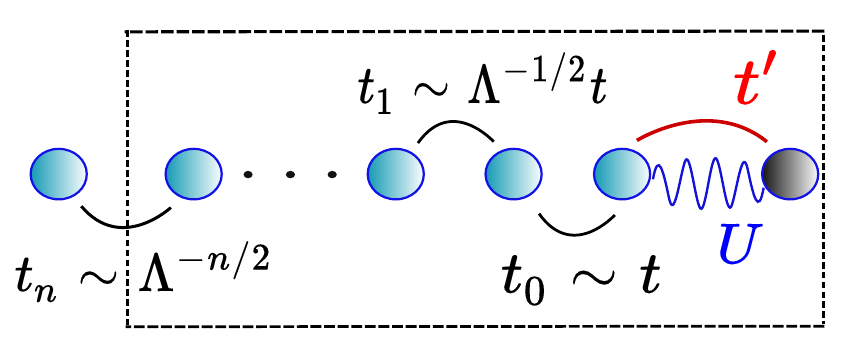}
\end{center}
\caption{Schematic NRG approach for the IRLM. Adding a new site to the existing chain of $N$ sites + impurity enlarges the Hilbert space at lower energy scales, from which only the lowest $S_{K}$ energy states are kept after truncation. }
\label{figure_model}
\end{figure}

The NRG method was originally developed by Wilson~\cite{WilsonNRG}. The method is quite standard, with many good reviews existing in the literature; here we refer the reader to the literature~\cite{Costi} for details on the method. For convenience we used a tight binding model for the lead; in the ideal case the lead is uniform $t_{n}\equiv t$ and semi-infinite $N\rightarrow \infty$. Numerics are performed using the Numerical Renormalisation Group (NRG)~\cite{WilsonNRG} which has finite $N$ and introduces a logarithmic discretization of the lead band by taking hopping amplitudes $t_{n}$ depend on $\Lambda$ as $t_{n}\sim \Lambda^{-n/2}$ where $\Lambda$ is the discretisation parameter; the uniform lead limit corresponds to $\Lambda\to 1$. This logarithmic discretization of the band is suitable to capture all low-energy properties of the original tight binding hamiltonian. The main property of the lead relevant for this work is the density of states at the Fermi level $\nu\sim t^{-1}$, which corresponds to a bandwidth of $2t$ in the uniform lead. Throughout this work, we have focused on the repulsive regime $U\geq 0$ of the interaction.

In order to compute the spectral function given by Eq.~\eqref{eq_A}, a broadening procedure must be chosen when computing the Lehmann representation of $A(\omega)$, which is given by:
\begin{multline}
A(\omega)=\frac{1}{Z(0)}\sum_{n}|\langle GS|d^{\dagger}|n\rangle|^{2}\delta(\omega- E_{n})\\
 + |\langle n|d^{\dagger}|GS\rangle|^{2}\delta(\omega+ E_{n})
\end{multline}
where $n$ labels the eigenstates, $|GS\rangle$ refers to the zero temperature ground state of the model, and $Z(0)$ is the partition function at $T=0$. 

\begin{figure}[!h]
\begin{center}
\includegraphics[clip=true,scale=0.8]{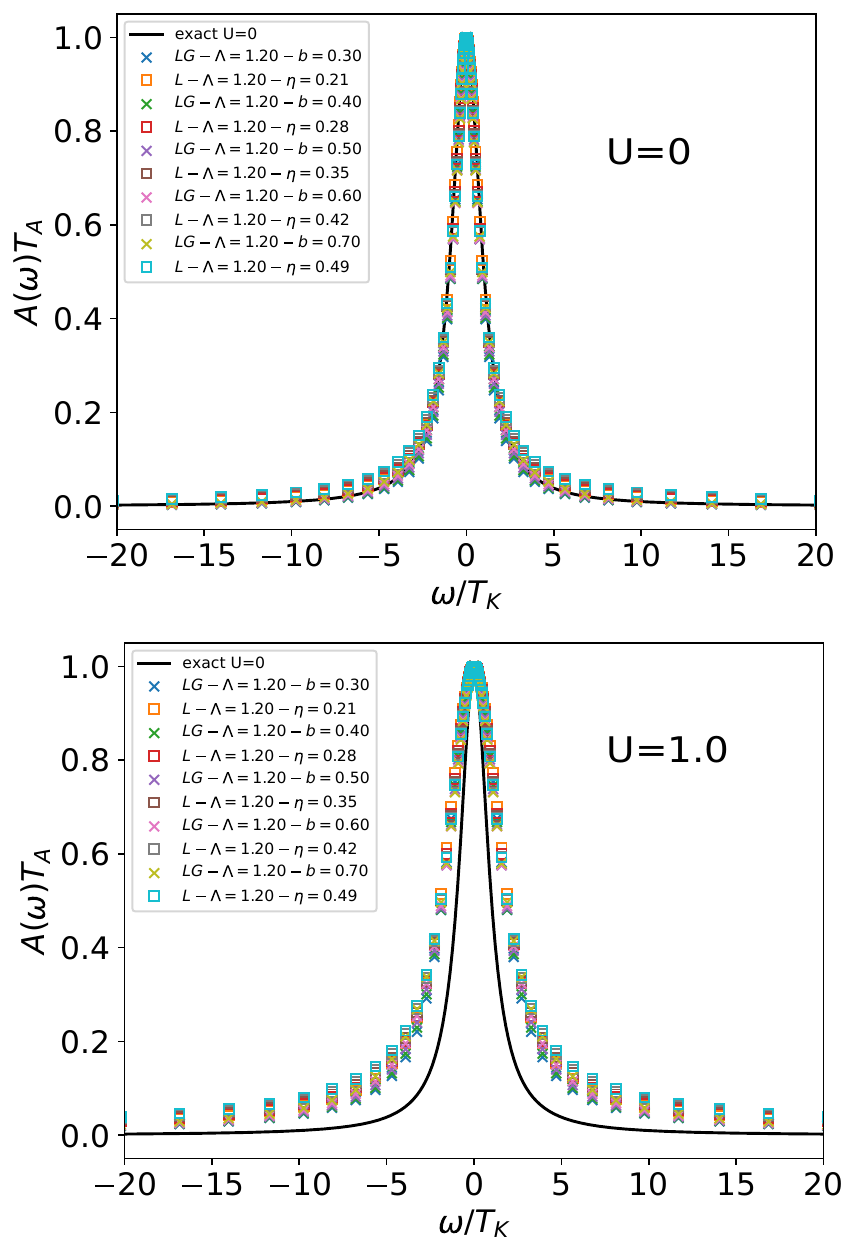}
\end{center}
\caption{Scaled figures showing convergence of numerical results for different values of the broadening parameters $b,\eta$ employed in the different broadening procedures ($LG$ stands for log-Gaussian and $L$ for Lorentzian) used in this work. The values of $U=0,1.0$ have been used (corresponding to $g=0.0$ and $g\sim 0.295$), which corresponds to a region before the central splitting occurs. The values of $\Lambda=1.2$, $t'=0.05$ and $S_{K}=1200$ were employed, on a lattice with $N=120$ sites. }
\label{figure_convergence_1}
\end{figure}

\begin{figure}[!h]
\begin{center}
\includegraphics[clip=true,scale=0.8]{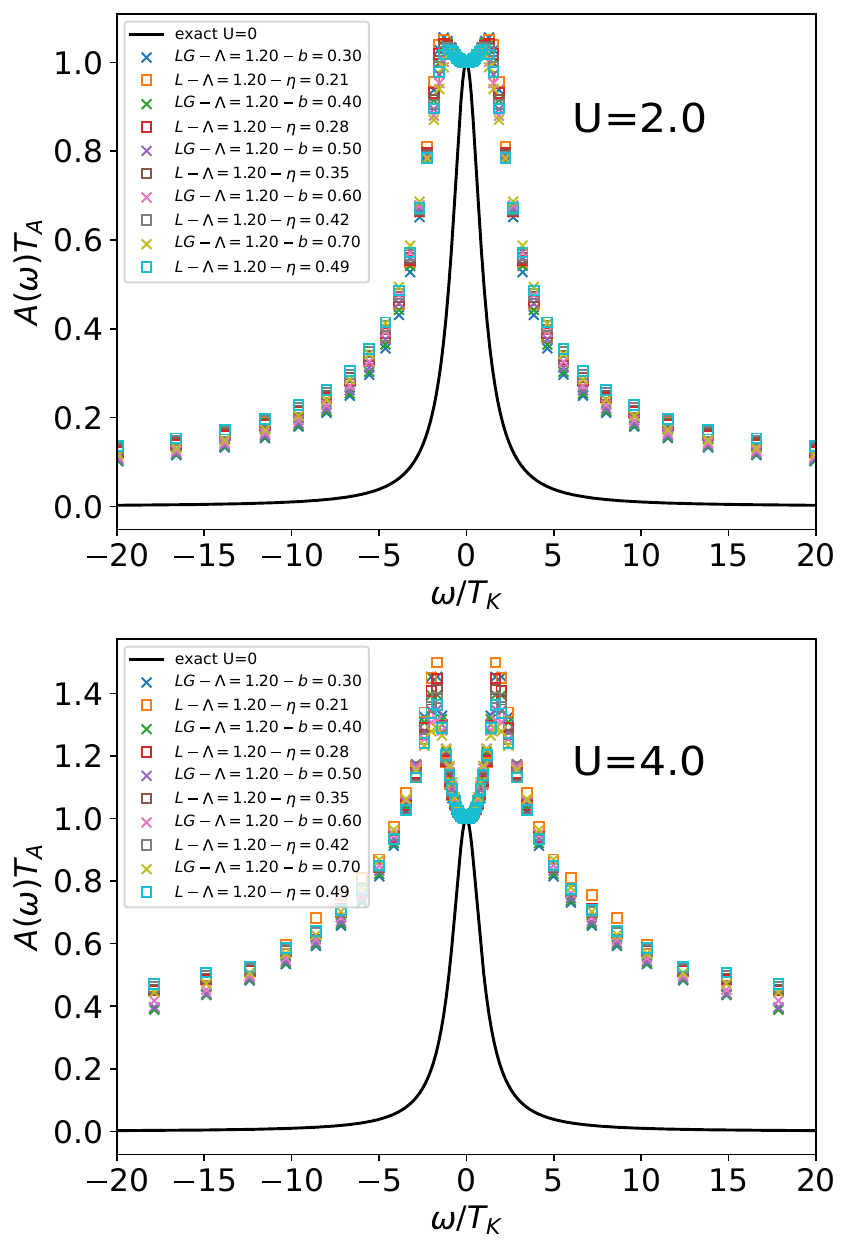}
\end{center}
\caption{Scaled figures showing convergence of numerical results for different values of the broadening parameters $b,\eta$ employed in the different broadening procedures ($LG$ stands for log-Gaussian and $L$ for Lorentzian) used in this work. The values of $U=2.0,4.0$ have been used (corresponding to $g=0.5$ and $g\sim 0.7$), which corresponds to a region after the central splitting occurs. The values of $\Lambda=1.2$, $t'=0.05$ and $S_{K}=1200$ were employed, on a lattice with $N=120$ sites. }
\label{figure_convergence_2}
\end{figure}

The delta functions are broadened to give smooth results. Two common broadening procedures are the log-Gaussian~\cite{Costi} and the Lorentzian broadenings, which were used in this work. For the log-Gaussian we used:
\begin{eqnarray}
\text{LG}:\hspace{10pt} P_{\text{LG}}(\omega\pm E_{n})= \frac{e^{-b^{2}/4}}{bE_{n}\sqrt{2\pi}}e^{-\frac{(\ln(|\omega|/E_{n}))^{2}}{b^{2}}}
\end{eqnarray}
with $b\in[0.3,0.7]$. For the Lorentzian broadening, the parameter $\eta=b/\sqrt{2}$ specifies the broadening of the Lehmann representation in Eq.~\eqref{eq_A}. In the main text, the Lorentzian broadening was used to resolve the Hubbard bands in Fig.~\ref{figure0}. For the rest of calculations, it is more convenient to employ the log-Gaussian broadening~\cite{Costi}.

We attach a convergence test in terms of the $b,\eta$ parameters used in both broadening procedures. We fix $t'=0.05$, $\Lambda=1.2$ and $S_{K}=1200$ for the total number of kept states, while we use a chain of $N=120$ sites. The main results do not differ in qualitative terms when one stays within the mentioned range of $b$. In figures like Fig.~\ref{figure_2}, a value $b\sim 0.6$ was employed.

\section{Diagrammatic Renormalisation Group}\label{app_B}
Here we show the derivation of equation~(\ref{gamma_exp}) using diagrammatic renormalisation group techniques. Let us write the spectral function in a general form:
\begin{equation}
A(\omega) = \frac{1}{\pi} \frac{h(\omega) T(\omega)}{\omega^2+T^2(\omega)}.
\label{Arenorm}
\end{equation}
At energy scales of the bandwidth, $\omega\sim D$, we have $h(D)=1$ and $T(D)=T_0$, there bare values. According to the renormalisation group paradigm, in the presence of interactions these parameters will flow as energy is lowered, with $T$ being the (effective) renormalised resonance width, while $h$ is known as the multiplicative renormalisation factor.\\ 

These flow equations were derived 40 years ago by Schlottmann~\cite{Schlottman}, and are given by:
\begin{equation}
\frac{dh}{d|\omega|} = g^2 \frac{h}{|\omega|+T},\;\;\;\;\; \frac{dT}{d|\omega|} = (-2g+g^2) \frac{T}{|\omega|+T}.
\end{equation}
Our variables $g,h,T$ correspond in Schlottmann's notation to $\gamma\rho,d,\Omega$.  Schlottmann also gives the flow equation for the vertex function $\Gamma$, however we have trivially eliminated this from the other two equations using the Ward identity he derives $d(\Gamma d)/d\omega=0$. Integrating the latter of these for $\omega \gg T$ gives $T/T_0 = (\omega/D)^{-2g+g^2}$. The essence of the renormalisation analysis is that the flow of the resonance width should be cut off when $T(\omega) \sim \omega$ which defines $T_K$.  Running through the calculation gives $T_K \sim T_0^{\alpha/2}$ with the exponent $\alpha$ as defined above.  Similarly, integrating $h$ from the bandwidth down to $T_K$ where the flow is cut off gives $h \sim T_K^{g^2}$, which gives $A(0) = T_K^{g^2-1}$ agreeing with Eq.~\eqref{gamma_exp} that was demonstrated numerically.

\bibliography{PAPER_BIBLIOGRAPHY}

\end{document}